# Applying Dynamic Training-Subset Selection Methods Using Genetic Programming for Forecasting Implied Volatility

Sana Ben Hamida (sana.ben_hamida@u-paris10.fr)
Paris West University, Department of Mathematics and Computer Science, NANTERRE, FRANCE

Wafa Abdelmalek (wafa.abdelmalek@fsegs.rnu.tn)
Research Unit MODESFI, Faculty of Economics and Business, SFAX, TUNISIA

Fathi Abid (fathi.abid@fsegs.rnu.tn)
Research Unit MODESFI, Faculty of Economics and Business, SFAX, TUNISIA

**Abstract:**
*Volatility is a key variable in option pricing, trading and hedging strategies. The purpose of this paper is to improve the accuracy of forecasting implied volatility using an extension of genetic programming (GP) by means of dynamic training-subset selection methods. These methods manipulate the training data in order to improve the out-of-sample patterns fitting. When applied with the static subset selection method using a single training data sample, GP could generate forecasting models which are not adapted to some out-of-sample fitness cases. In order to improve the predictive accuracy of generated GP patterns, dynamic subset selection methods are introduced to the GP algorithm allowing a regular change of the training sample during evolution. Four dynamic training-subset selection methods are proposed based on random, sequential or adaptive subset selection. The latest approach uses an adaptive subset weight measuring the sample difficulty according to the fitness cases' errors. Using real data from S&P500 index options, these techniques are compared to the static subset selection method. Based on MSE total and percentage of non fitted observations, results show that the dynamic approach improves the forecasting performance of the generated GP models, especially those obtained from the adaptive-random training subset selection method applied to the whole set of training samples.*

## I. INTRODUCTION

Financial market volatility is a key variable in financial investment decisions and plays a central role in derivative valuation and in conducting dynamic hedging strategies. To assess the fair value of an option or to hedge market risk, an investor needs to specify his expectations regarding future volatility. Due to their forward-looking nature, option prices are especially useful for extracting such information. A number of investigations supported the idea of using implied volatility as a good predictor of future volatility (Latané and Rendelman (1976), Chiras and Manaster (1978), Fleming (1993), Blair et al. (2001), Corrado and Miller (2005)). Assuming that an option pricing model correctly represents investors' behavior, the implied volatility can be derived from observed option prices by appropriately inverting the option pricing model. In contrast, Genetic Programming (GP) offers explicit formulas which can compute directly the implied volatility expressed as a function of option prices and other observable variables. This volatility's forecasting approach should be free of strong assumptions and more flexible than parametric models. GP (Koza (1992)) is an evolutionary-based search technique which is based on the principles of natural evolution. Using its basic and flexible tree-structured representation, GP is capable of solving some difficult problems without requiring the user to know or specify the form or the structure of the solution in advance.

GP has proved successful at forecasting time series volatility in different markets, such as foreign exchange and index markets. Neely and Weller (2002) have tested the forecasting performance of GP for USD-DEM and USD-YEN daily exchange rates against that of the generalized autoregressive conditional heteroskedasticity (GARCH) model (Baillie et al. (1996)) and a related RiskMetrics volatility forecast over different time horizons. According to various accuracy criteria, GP has produced significantly superior results. Using high frequency foreign exchange USD-CHF and USD-JPY time series, Zumbach et al. (2001) have compared the GP forecasting accuracy to that of historical volatilities and some popular autoregressive conditional heteroskedasticity (ARCH) type models, notably the generalized ARCH (GARCH (1,1)) model of Bollerslev (1986), the fractionally integrated GARCH (FIGARCH) model of Baillie et al. (1996) and the heterogeneous ARCH (HARCH) model of Müller et al. (1997). According to the root-mean squared errors, the generated GP volatility models have outperformed the benchmarks. Using historical returns of Nikkei 225 and S&P500 indices, Chen and Yeh (1997) have applied a recursive genetic programming (RGP) approach to estimate volatility by simultaneously detecting and adapting to structural changes. Results have shown that RGP is a promising tool for the study of structural changes. Applying a combination of techniques such as evolutionary algorithms GA and GP, Ma et al. (2006, 2007) have proposed a systematic approach to address specifically nonlinear problems in the forecast of financial indices using intraday data of S&P100 and S&P500 indices. As a result, accuracy of forecasting has reached an average of over 75% surpassing other publicly available results on the forecast of any financial index. Abdelmalek et al. (2009) have extended the studies mentioned earlier by forecasting the implied volatility of Black-Scholes from the S&P500 index call options instead of the integrated volatility based on historical returns. They have considered the problem of managing too large databases when training GP. As a consequence, they have proposed to split data into smaller subsets by time series and moneyness-time to maturity classes and to train GP separately on all learning sub-samples. The proposed approach is called static training-subset selection method. According to total and out-of-sample mean squared errors (MSE), results have shown that time series models seem to be more accurate in forecasting implied volatility than moneyness-time to maturity models. Such an approach has provided some local solutions not adaptive to the enlarged data set, especially when learning with a moneyness-time to maturity sample. According to Gilli (2010), the relationship between in-sample fit and out-of-sample performance is not monotonous and an optimal in-sample solution might be ineffective when applied to out-of-sample data.

The present paper investigates the application of a dynamic subset selection method, in which the training subset samples change during the GP run. This allows GP to learn simultaneously on all training sub-samples rather than just a single subset, which seems to have better generalization ability. This technique aims to intensify search space exploration and thus enhance the robustness of GP with large data set. Robustness is an important feature of an evolved program (Ito et al. (1996)). It is defined as the ability to cope with noisy or unknown situations. The use of the dynamic training-subset selection method could reduce the problem of fitting out-of-sample patterns and could improve the forecasting accuracy. The major contribution of this paper is the use of the *Adaptive Subset Selection* method (*ASS)* which is performed in proportion to a ratio of difficulty associated to each training sample. Two other selection methods are used for purposes of comparison: the *Random Subset Selection* method (*RSS*) where samples are selected in a random way and the *Sequential Subset Selection* method (*SSS*) where samples are selected in a regular way. Comparative experiments are provided to show how dynamic training subset-selection methods are applied to improve the robustness of GP to generate general models relative to static training-subset selection method. Using Total MSE and percentage of non fitted observations (NFO) as performance criteria, results show that the forecasting accuracy is improved with the *Adaptive Subset Selection* method (*ASS)*.

The remainder of the paper is organized as follows. Section 2 presents the research objectives and theoretical foundation regarding forecasting implied volatility and usefulness of subset selection methods. Section 3 illustrates the research design and methodology. Section 4 provides a description of the dynamic subset selection methods implemented. Section 5 reports and discusses the results of the comparison between static and dynamic selection methods. Finally, section 6 summarizes and gives perspectives.

## II. RESEARCH OBJECTIVES AND THEORETICAL FOUNDATION

This paper addresses the application of GP by means of a dynamic subset selection method for the purposes of forecasting implied volatility. The objective is to improve forecasting accuracy.
In financial volatility forecasting, it is important to find accurate models fitting a maximum number of input cases from learning data. The goal is to correctly predict the volatility of new input data. By considering the assumption that there are inherently many different patterns in financial series (Povinelli (1999)), instead of using one formula (model) to explain the entire data, a better idea would be to select a set of best models that could be combined and used to forecast future implied volatility values. As a consequence, the main interest of this work lies in applying a dynamic subset selection method for training GP on the full training samples, so that the entire input data become in-learning-sample. The main contribution of this paper is the use of the adaptive subset selection method. This method is inspired by the dynamic subset selection proposed by Gathercole and Ross (1994). It is based on the assumption that there is a benefit in focusing the GP's abilities on difficult training samples, i.e., the ones that have the highest MSE. The challenge is to make GP adaptive to all training samples and be able to generate general models. The originality of this method is to assign weights to each training subset and update these weights through the generations. Initial weight is assigned according to the initialization method adopted (random or sequential) and increases each time an individual is not able to solve the corresponding fitness sample cases. This approach lightens the training task for GP and favors the discovery of solutions that are more robust across different data samples.

In the following sub-sections, background information regarding implied volatility is first introduced. Second, a discussion about the usefulness of subset selection methods is made. This is to explain the originality of the dynamic selection methods created versus the use of previous sampling methods.

### 2.1. Implied volatility

There are two approaches to generate volatility forecasts. One is to extract information about the variance of future returns from their history; the second is to elicit market expectations about the future volatility from observed option prices. Options markets provide market participants and policy-makers with a rich source of information for gauging market sentiment. An option contract is a derivative security that gives the holder the right to buy (call) or to sell (put) the underlying asset by a certain date for a certain price. The price in the contract is known as the exercise price or strike price. The date in the contract is known as the expiration date or maturity. American options can be exercised at any time up to the expiration date. European options can be exercised only on the expiration date itself. Options contracts can be divided into several classes according to either moneyness or term to expiration. By the term to expiration, an option contract can be short-term (ST), medium-term (MT) and long-term (LT). By the moneyness, a call option can be in-the-money (ITM) if the stock price is above the strike price (S>K), out-of-the-money (OTM) if the stock price is below the strike price (S<K) and vice versa for a put option. If the strike price is closest to the current value of the underlying stock (S=K), the option contract is said to be at- the- money (ATM). The difference between the stock price and the strike price represents the intrinsic value for a call option. It can be a positive number, or zero otherwise. The total value of an option called premium is basically the sum of its intrinsic value and its time value.

In the Black-Scholes (BS) framework, the option price is a function of variables which are directly observable except for the volatility. The price of an option therefore depends on the market's opinion about the future volatility of the underlying asset upon which the option is written.
The Black-Scholes (1973) option pricing model assumes that volatility is constant. It was first derived for the European call option written on a non-dividend paying stock, as defined in equation (1).

$$C_{BS} = S\Phi(d_1) - Ke^{-r\tau}\Phi(d_2) \qquad (1)$$

Where, $d_1 = \dfrac{\ln\left(\dfrac{S}{K}\right) + (r + 0.5\sigma^2)\tau}{\sigma\sqrt{\tau}}, d_2 = d_1 - \sigma\sqrt{\tau}$.

$C_{BS}$ denotes the price of a European call option, S is the market price of the underlying asset, K is the strike price of the option, r is the risk-free interest rate, $\tau$ is the time to maturity, $\Phi$ is the cumulative normal distribution function and σ is the volatility.

By equating the observed market price $C_t^*$ of an option with the BS price $C_{BS}$ and implicitly solving for σ, an implied volatility can be found.

$$\exists! \sigma_t^{BS}(K,T) \succ 0,$$
$$C_{BS}(S_t, K, \tau, \sigma_t^{BS}(K,T)) = C_t^*(K,T) \quad (2)$$

According to the BS assumptions, this implicitly calculated volatility should be constant. However, it can be easily shown empirically that the implied volatility is not constant and changes with different option strike prices and expiry dates. For example, short-dated options will be less sensitive to implied volatility, while long-dated options will be more sensitive. This is based on the fact that long-dated options have more time value priced into them, while short-dated options have less. Besides, options with strike prices that are near the money are most sensitive to implied volatility changes, while options that are further in-the-money or out-of-the-money will be less sensitive to implied volatility changes.

Implied volatility represents the expected volatility of a stock over the life of the option. As expectations change, option premiums react appropriately. Implied volatility is directly influenced by the supply and demand of the underlying options and by the market's expectation of the share price's direction. As expectations rise, or as the demand for an option increases, implied volatility will rise. Options that have high levels of implied volatility will result in high-priced option premiums. Conversely, as the market's expectations decrease, or demand for an option diminishes, implied volatility will decrease. Options containing lower levels of implied volatility will result in cheaper option prices. This is important because the rise and fall of implied volatility will determine how expensive or cheap time value is to the option.

### 2.2. Subset selection methods: a discussion

Evolving programs is often a time consuming task, in particular in terms of fitness evaluation's effort. When using GP with a large set of training cases and a large population size, a very large number of tree evaluations must be carried out every generation. Many methods try to reduce the number of such evaluations by selecting a small subset of the training data set during fitness evaluation. These methods differ in how they choose proper subsets from the set of all fitness cases[1] for evaluation. The simplest technique is to use a static subset. However, using a single learning sample might lead to a local optimum that solves only a part of the fitness cases. To reinforce learning from different parts of the search space, some solutions were proposed such as carrying out several runs using one subset for each run and selecting the appropriate model from different resulting models (Abdelmalek et al. 2009). The *Historical subset selection* (Gathercole and Ross, 1994) extends the static subset selection by recording all fitness cases that are not solved by the best population member in any given generation over a small number of runs. These fitness cases become part of a static subset and are used in further GP runs.

A more flexible method is to pick a variety of subsets during the course of a training run. There are many ways to select different subsets from the training set. The goal is to pick the right subsets to allow the learning algorithm to proceed as fast and as accurately as possible. The simplest method for picking a different set for each generation is random. *Random subset selection* (Gathercole and Ross, 1994) chooses a new subset for each generation. Each learning data instance is selected independently with equal probability, which leads to varying subset sizes. *Stochastic sampling* (Nordin and Banzhaf, 1997; Banzhaf et al., 1998) chooses a new subset for each generation and for each individual, respectively, where all data cases having the same probability of being selected. As a result, different individuals will probably be evaluated on different data samples, which cast some doubts on the fairness of the selection step in the evolutionary algorithm (section 3.2).

More efficient criteria can be used to guide the selection of the new subset based on fitness-case topology or the performance of the current GP population. Fitness-case-topology based sampling (Lasarczyk and al., 2004) relies on creating a dynamic weight for each couple of fitness case that is updated by the evolutionary system according to the number of GP solutions able to solve both of the fitness cases of each couple. The learning subset is constructed by a random selection from the set of couples with small weights. *Dynamic subset selection* (Gathercole and Ross, 1994; Gathercole and

---

[1] A fitness case is an input/output pair, which measures how well an evolved individual predicts the output(s) from the input(s).

Ross, 1997; Gathercole, 1998) makes use of the difficulty of each training case, i.e. how often it is misclassified, and its age, i.e. how many generations since it was last selected. This has worked well on some large classification problems, using less computer resources to produce better results than standard GP.

However, the re-sampling procedure during the course of a training run increases the complexity of the GP system and requires the entire data set to be stored in the main memory. With GP systems, using the full learning data might be impossible when the input sample does not fit within the main memory and could cause serious problems in the realization of predictors. In this case, data reduction through the partitioning of the data set into smaller subsets seems to be a good approach (Abdelmalek et al. (2009)).

In this paper, we suggest to construct subset data with a fixed size from the full database using data division schemes and to apply dynamic training-subset selection methods to select subsets already built up for learning process. With dynamic subset selection methods, a new subset is picked each g generations (g is the number of generations to change sample). These methods can be used to prevent a biasing influence of subset selection in evolution. These methods differ from the existing approaches as they don't extract a fixed number of fitness cases from the training set, but select a subset from a set of sub-samples data already built up using the data division scheme, which avoids increasing the complexity of the GP engine. In this paper, we proposed four dynamic training-subset selection methods: *Random Subset Selection* method (*RSS*), *Sequential Subset Selection* method (*SSS*), *Adaptive-Sequential Subset Selection* method (ASSS) and *Adaptive-Random Subset Selection* method (ARSS). The *RSS* and *SSS* allow the genetic programming to learn on all training samples sequentially (*SSS*) or randomly (*RSS*). However, with these methods, there is no certainty that GP will focus on the samples which are difficult to learn. Then, the *ASSS* and the *ARSS,* which are variants of *the adaptive subset selection (ASS),* are introduced to focus the genetic programming's attention onto the difficult samples i.e. having the greatest MSE and then to improve the learning algorithm. An adaptive weight is associated to each subset (Adaptive Subset Weight) and updated each generation according to the average fitness of the all cases in the corresponding sample.

### III. RESEARCH DESIGN AND METHODOLOGY

To achieve the research goals presented above, we followed three steps. The initial stage is devoted to the data preparation, to the implementation of GP[2] and to the subset selection procedures. The second step is devoted to the learning process using static and dynamic subset selection methods. The last step is dedicated to results comparison and the selection of the best forecasting models. The following research steps are summarized in the scheme below.

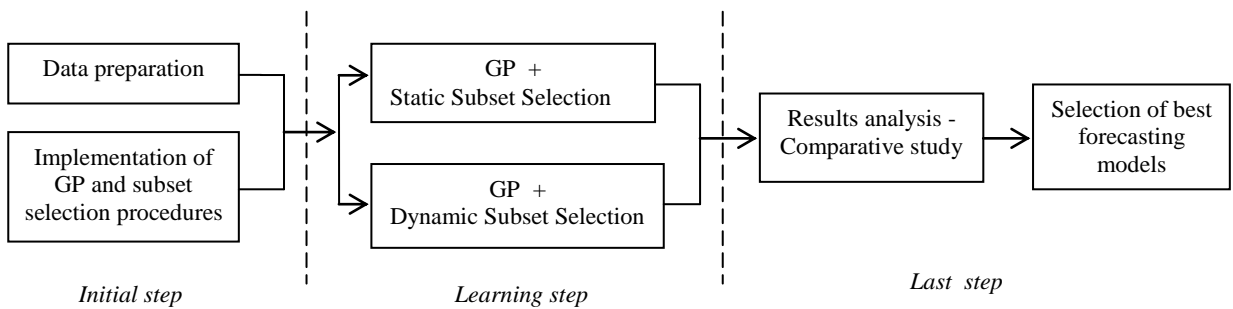

### 3.1. Data preparation

The data used to perform our empirical analysis are daily prices of European S&P500 index calls options, from the Chicago Board of Options Exchange (CBOE), for the sample period running from January 2, 2003 to August 29, 2003. The database includes the time of the quote, the expiration date, the exercise price and the option price. Similar information for the underlying S&P 500 index is also available on a daily basis. The daily bid and ask quotes for the call options are obtained from the CBOE. Following a standard practice, we use the average of an option's bid and ask price as a stand-in for the market value of the option. Strike price intervals are 5 points, and 25 for far months. The expiration months, are three near term months followed by three additional months from the March quarterly cycle (March, June, September, and December). The risk free interest rate is approximated by using 3 month US Treasury bill rates.

---
[2] GP system is built around the Evolving Object library, which is an *ANSI-C++ evolutionary computation Framework* (EO library).

Two preparation procedures have been applied to the data before use, preprocessing and division. The data preprocessing serves the purpose of "smoothing" the raw data and removing what is not essential, before the machine learning algorithm is applied. It is widely accepted that preprocessing is usually beneficial and has positive effects on the learning process (Chen et al. (2007)).

The original training set contains 42504 daily observations of call option prices and their determinants. To reduce the likelihood of errors, data screening procedures are used (Harvey and Whaley (1991, 1992)). Then, four exclusion filters are applied to construct the final sample. First, call options with time to maturity less than 10 days are excluded from the sample. This can be explained by the fact that implied volatilities of short-term options are very sensitive to small errors in the call price and may convey liquidity-related biases. Second, call options with low quotes are eliminated to mitigate the impact of price discreteness on option valuation. Third, deep-ITM and deep-OTM option prices are also excluded due to the lack of trading volume. Finally, option prices not satisfying the arbitrage restriction (Merton (1973)), $C \geq S - Ke^{-r\tau}$, are not included. The final sample contains 6670 daily option quotes, with at-the-money (ATM), in-the-money (ITM) and out-of-the-money (OTM) options respectively taking up 37%, 34% and 29% of the total sample.

For the data sub-sampling procedure, two schemes were used. For the first division scheme, the full sample is sorted by time series (TS), and for the second, by moneyness-time to maturity (MTM). For time series, data are divided chronologically into successive samples ($S_1$, $S_2$… $S_{10}$), each containing 667 daily observations. These samples will be used simultaneously for training and test steps. For moneyness-time to maturity, data are divided into nine classes with respect to moneyness and time to maturity[3]. Each class $C_i$ is divided into a training set $C_i^L$ and a test set $C_i^T$, which produces respectively nine training and nine test moneyness- time to maturity sub-classes. Figure 1 illustrates the two division schemes.

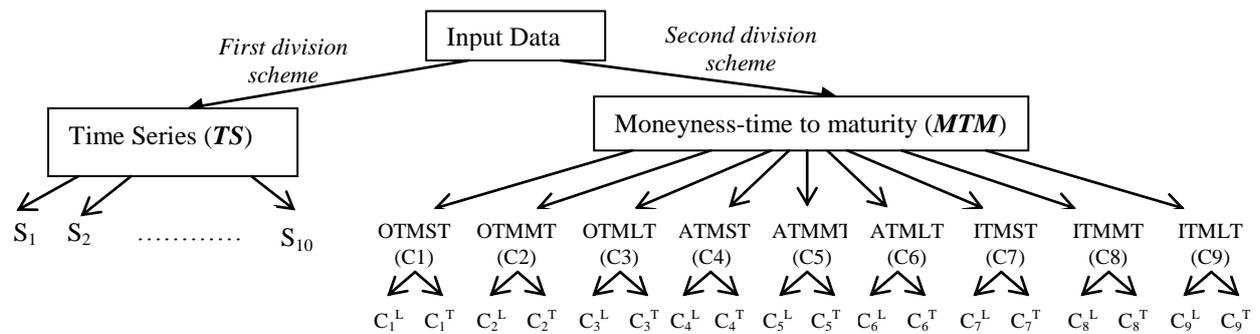

**Figure 1**: Data division schemes

### 3.2. The design of genetic programming

There are several GP techniques that might be used to deal with financial forecasting. People are familiar with regression analysis. One problem with regression analysis is that the results of the analysis depend very much on the skills and inventiveness of the experimenter. Furthermore, in many application areas there is a considerable tradition of using only linear or quadratic models, even when the data might be better fitted by a more complex model. Symbolic regression attempts to go beyond this. It consists in finding a function that fits some target values without making any assumptions about the structure of that function. Symbolic regression was one of the earliest applications of GP (Koza, 1992), and has continued to be widely studied (Cai et al. (2006); Gustafson et al. (2005); Keijzer (2004); Lew et al. (2006)). The steps necessary to implement the GP's symbolic regression are summarized in algorithm 1.

---

[3] According to moneyness criterion: A call option is said out-of-the money (OTM) if $S/K < 0.98$; at-the-money (ATM) if $S/K \in [0.98, 1.03[$; and in-the-money (ITM) if $S/K \geq 1.03$. According to time to maturity criterion: A call option is Short Term (ST) if $\tau < 60$ days; Medium Term (MT) if $\tau \in [60, 180]$ days; and Long Term (LT) if $\tau > 180$ days.

*Initialize population*
***while*** *(termination condition not satisfied)* ***do***
***begin***
    *Evaluate the performance of each individual according to the fitness criterion*
    ***Until*** *the offspring population is fully populated* ***do***
        *- Select individuals in the population using the selection algorithm*
        *- Perform crossover and mutation operations on the selected individuals*
        *- Insert new individuals in the offspring population*
    *Replace the existing population by the new population*
***end while***
*Report the best solution found*
***end***

**Algorithm 1**: Pseudo code of Genetic Programming.

*Terminal and function sets*

The standard GP tree is a simple structure, made by terminal (or leaf) nodes, and non-terminal (or function) nodes with branches. Terminal and function sets, which are described in Table 1, define the ingredients that GP can use to create function models and to construct potential solutions.

| **Expression** | | **Definition** |
|---|---|---|
| Terminal Set | C/K | Call price / Strike price |
| | S/K | Index price / Strike price |
| | $\tau$ | Time to maturity |
| Function Set | + | Addition |
| | - | Subtraction |
| | * | Multiplication |
| | % | Protected division: x % y = 1 if y=0;   x % y =  x % y otherwise |
| | cos | Cosinus function |
| | Sin | Sinus function |
| | ln | Protected natural log: $\ln(x) = \ln(|x|)$ |
| | Exp | Exponential function: $\exp(x) = e^x$ |
| | Sqrt | Protected square root: $\sqrt{x} = \sqrt{|x|}$ |
| | Ncdf | Normal cumulative distribution function $\Phi$ |

**Table 1:** Terminal set and function set.

The terminal set includes input variables, mainly, the call option price divided by strike price $\frac{C}{K}$, the index price divided by strike price $\frac{S}{K}$ and time to maturity $\tau$. The predictive target output is the implied volatility $\sigma_t^{BS}$ computed using the Black-Scholes formula. The function set includes basic mathematical operators and Black Scholes components. The mathematical operators we use are the basic arithmetic operators together with the cosine functions and the sine functions. The Black Scholes components involve the log function (ln), the exponential function (exp), the square root function ($\sqrt{\phantom{x}}$) and the normal cumulative distribution function ($\Phi$), which may be useful for implied volatility models.

*Initialization*

GP starts by randomly creating an initial population of trees, which are generated by randomly picking nodes from a given terminal set and function set (table 1). The initialization scheme used in this paper is the ramped half-and-half method (Koza (1992)), which is a combination of the full and grow initialization methods. This method involves generating an equal number of trees using a maximum initial depth that ranges from 2 to 6, as specified in Table 2. For each level of depth, half of initial trees are generated via the full method, and the other half is generated via the grow method.

*Fitness function*

The evolutionary process is driven by a fitness function that evaluates the performance of each individual (tree) in the population. The fitness criterion used for the selection of the best individuals is the mean squared error (MSE) between the target output volatility ($\sigma_t^{BS}$) and the generated GP volatility ($\hat{\sigma}_t$), computed as follows:

$$\text{MSE} = \frac{1}{N} \sum_{t=1}^{N} \left( \sigma_t^{BS} - \hat{\sigma}_t \right)^2 \quad (3)$$

Where, N is the number of fitness cases in the learning sample. At the end of evolution, each individual is evaluated according to the MSE computed with the equation (3) using a test sample which must be different from the learning sample.

*Selection*

Based on fitness measure, GP probabilistically selects the fitter individuals from the population to act as the parents of the next generation. Selection determines which individuals of the population will have all or some of their genetic material passed to the next generation. In general, parents displaying a higher level of performance are more likely to be selected with the hope that they can produce better offspring with larger chance. The most commonly used method for selecting individuals in GP is tournament selection. In tournament selection, a number of individuals, called the tournament size, are selected randomly from the population and they compete with each other. The best is to be selected. As specified in Table 2, the tournament size used for experiments is equal to 4.

*Genetic operators*

Crossover and mutation are the two basic operators which are applied to the selected individuals in order to generate new individuals for the next generation. They are needed to explore the search space.

Crossover operator:

The most commonly used form of crossover is subtree crossover. Given two parents, subtree crossover randomly selects a crossover point (a node) in each parent tree. Then, it creates the offspring by replacing the subtree rooted at the crossover point in a copy of the first parent with a copy of the subtree rooted at the crossover point in the second parent. All non-terminal nodes (except the root node) have the same probability to be selected. The generated offspring should not surpass the fixed size. As indicated in Table 2, the crossover operator is used to generate about 60% of the individuals in the population.
The maximum tree size (measured by depth) allowed after the crossover is 17. This is a popular number used to limit the size of the tree (Koza (1992)). It is large enough to accommodate complicated formulas and works in practice.

Mutation operator:

The basic role of the mutation operator in the evolutionary process is to ensure diversity in the population. It affects small random changes in a tree by randomly altering nodes or sub-trees to create a new offspring and continue the search process. Many mutation operators are used in GP. The most commonly used form of mutation in GP is subtree (or branch) mutation. It replaces a randomly selected subtree with another randomly created subtree (Koza (1992)). Another common form of mutation is point mutation (or node replacement), which randomly changes a node in the individual and replaces it with another node with the same arity (McKay et al. (1995)). Expansion mutation randomly selects a terminal node in the tree, and then replaces it with a new randomly-generated subtree. As indicated in Table 2, Branch mutation is applied with a rate of 20%; Point and Expansion mutations are applied with a rate of 10% respectively.

The parameter choices for crossover and mutation are clearly critical in ensuring a successful GP application. They impact on populational diversity and the ability of GP to escape from local optima (Yin et al. (2007)).

### *Replacement*

Once the new population has been created, the current population (parents) is replaced by the new one (offspring). We use a comma replacement method to replace parents for the next generation (Schwefel (1995)). This method selects the best offspring to replace the parents. If μ is the population size and λ is the number of new individuals (which can be larger than μ), the population is constructed using the best μ out of the λ new individuals.

### *Termination criterion*

The termination criterion we use is the maximum number of generations to be run. We take 400 and 1000 for static and dynamic training-subset selection, respectively. In the dynamic training-subset selection approach, the maximum number of generations is increased to allow GP to train on the maximum of samples simultaneously. Typically, the single best-so-far individual is then obtained and designed as the result of the run.

Parameters of the GP design used in this work are summarized in Table 2. The optimal set of genetic parameters is determined based on a series of trial and error experiments.

| | |
|---|---|
| Population size: | 100 |
| Offspring size: | 200 |
| Generations' number for static method: | 400 |
| Generations' number for dynamic method: | 1000 |
| Generations' number to change sample | 20-100 |
| Maximum depth of new individuals: | 6 |
| Maximum depth of the tree: | 17 |
| Tournament size: | 4 |
| Crossover probability: | 60% |
| Mutation probability: | 40% |
|     Branch mutation: | 20% |
|     Point mutation: | 10% |
|     Expansion mutation: | 10% |

**Table 2 :** Summary of GP parameters.

### IV. DESCRIPTION OF DYNAMIC TRAINING-SUBSET SELECTION METHODS

To implement the dynamic training-subset selection methods, two supporting decision-designs are necessary to be fixed. First, the frequency of the replacement of the current training subset; such a frequency is designed by a fixed iteration number (g) added as a parameter to the GP system. The value of this parameter must be chosen such as the GP system has enough time to adapt the genetic material in the population to make it able to solve the current subset. However, g does not have a very high value in order to keep the population diversity necessary for the next learning step.

Second, the procedure of replacement and selection of the new training sample is determined. To design the subset replacement protocol, two approaches are available. In the first approach, all sub-samples are treated equally. Sub-samples are then selected with a uniform probability (*Random Subset Selection* method or *RSS*) or in a regular way, by taking the samples in turn (*Sequential Subset Selection* method or *SSS*). In the second approach, subset selection is performed proportionally to a predefined ratio of difficulty (*Adaptive Subset Selection* method or *ASS*). It involves an *Adaptive*

*Subset Weight* updated at each g generations, according to the fitness of the best current model computed according to all observations in each subset. The lower the sample performance is, the higher the selection pressure is. Samples having a great weight are selected more frequently in the learning process. The goal of this approach is to focus on the samples having the fitness cases the most difficult to learn. The GP is then guided to adapt its models to the more difficult samples. Only the individuals with the desirable characteristics that are well adapted to the environmental change will survive. Thereby, the problem of over-fitting that can be encountered with the static subset selection can be prevented. We have two variants of ASS: *Adaptive-Sequential Subset Selection* method (ASSS) and *Adaptive-Random Subset Selection* method (*ARSS*), differing on the initialization procedure of the sample weights. The operating principle of the dynamic selection is as follows.

Let S be the set of training samples $S_{i\ (i=1...k)}$, where k is the total number of samples. A selection probability P ($S_i$) is assigned to each sample $S_i$ which is changed each g generation (g is the number of generations to change sample) according to the dynamic training-subset selection method used. Once a new training sample is selected, the best individuals are used as population for the next training samples. This procedure is repeated until the maximum number of generations is reached. This permits GP to adapt its generating process to changing data in response to feedback from the fitness function.

To decide the training subset to select with the ASS method, an Adaptive Subset Weight (ASW) is computed for each subset according to the MSE values obtained along the last g generations. The ASW and the four proposed methods, RSS, SSS, ASSS and ARSSS, are described in the following.

### 4.1. Adaptive Subset Weight

For Adaptive Subset Selection method, the selection probability depends on the subset weights computed proportionally to the sample's average fitness. After g generations, the weight of the learning sample $S_i$ is updated as follows:

$$W(S_i) = \frac{\sum_{t=1}^{g} \sum_{j=1}^{M} f(X_j)}{M * g} \qquad (4)$$

Where M is the population size, g is the number of generations to change sample, and $f(X_j)$ is the MSE of the individual $X_j$, where $X_j \in P_t$ ($P_t$ is the current population).

If several individuals in the population have difficulty to solve some fitness cases in a sample $S_i$, then this sample will have a high subset weight and a great probability to be selected for the next learning step.

The update of the subset weights does not increase the complexity of the GP program and does not need additional computational cost as some dynamic subset selection methods yet proposed in the literature (Gathercole and Ross, 1997; Lasarczyk and al., 2004). Indeed, the individuals' fitness (used in equation (4)) are computed by the GP system and the subsets selection probabilities are computed only each g generation.

### 4.2. Random training-Subset Selection method (RSS):

This method randomly selects the training samples with replacement. All the samples from S have the same probability to be selected: P ($S_i$) =1/k, 1≤ i ≤ k. Figure 2 illustrates an example of the best fitness (MSE) curve along evolution using the RSS method. As selection of training samples is random, the performance of the current population changes with the training sample used for evolving the genetic program.

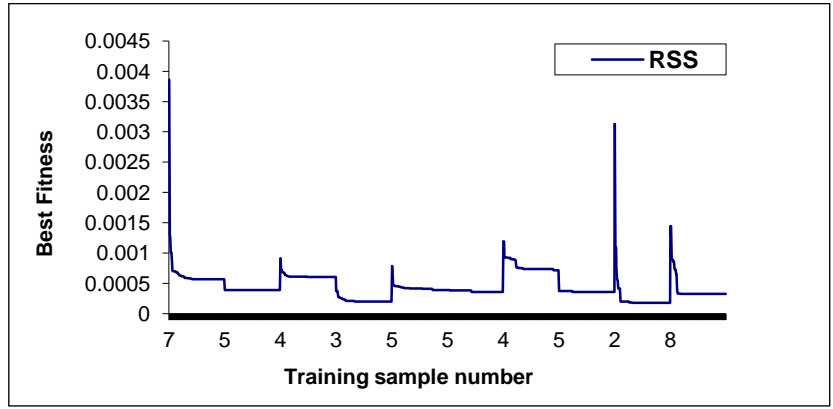

**Figure 2:** Example of fitness curve of the best individuals generated by GP using RSS method for TS samples

### 4.3. Sequential training-Subset Selection method (SSS)

This method selects all the training samples in turn. All the learning subsets are used during the evolution in an iterative way. If, at generation g-1, the current training sample is $S_i$, then at generation g: $P(S_j) = 1$, with j= i+1 if i<k, or j=1 if i=k. Figure 3 illustrates an example of the best fitness (MSE) curve along evolution using the SSS method. It shows that all the learning subsets are used during the evolution in an iterative way.

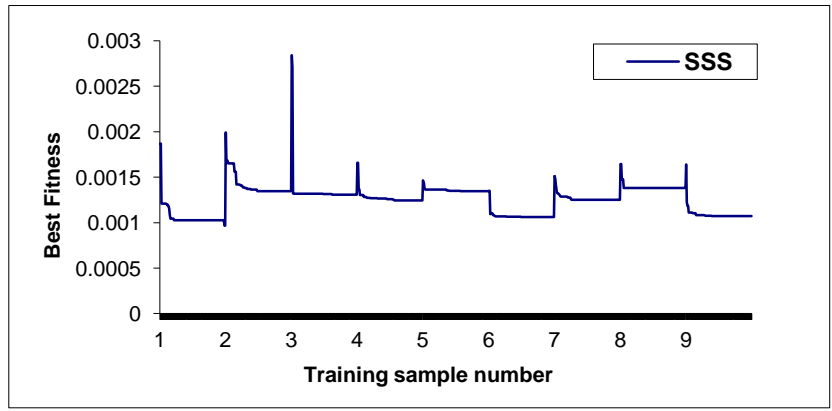

*Figure 3:* Example of fitness curve of the best individuals generated by GP using SSS method for MTM classes

### 4.4. Adaptive training-Subset Selection method (ASS):

Instead of selecting a training subset data in a random or sequential way, one can use an adaptive approach to dynamically select "difficult" training subset data having high fitness errors (MSE). This approach is inspired by the dynamic subset selection method of Gathercole and Ross (1994), which is based on the idea of dynamically selecting instances, not training samples, which are difficult and/or have not been selected for several generations. ASS simplifies this method by selecting training samples, not instances, containing unsolved fitness cases.

Selection is made according to the *Adaptive Subset Weight (ASW)*. After g generations, training samples are re-ordered, so that the most difficult training samples, those having higher *ASW*, will be moved to the beginning of the ordered training list, and the easiest training samples, those having smaller *ASW*, will be moved to the end of the ordered training list.

### a. Adaptive-Sequential training-Subset Selection method (ASSS):

The initial weights are initialized with a constant and the selection of samples is done in an iterative way: $W(S_i) = C, 1 \leq I \leq k$. Then, for the k first generations, selection of training samples is made in the order using the SSS method (subsection 4.1). Later, from the generation k+1, samples are selected for the next step according to the adaptive approach based on the re-ordering procedure (equation (4)). Figure 4 illustrates an example of the best fitness (MSE) curve along evolution using the ASSS method.

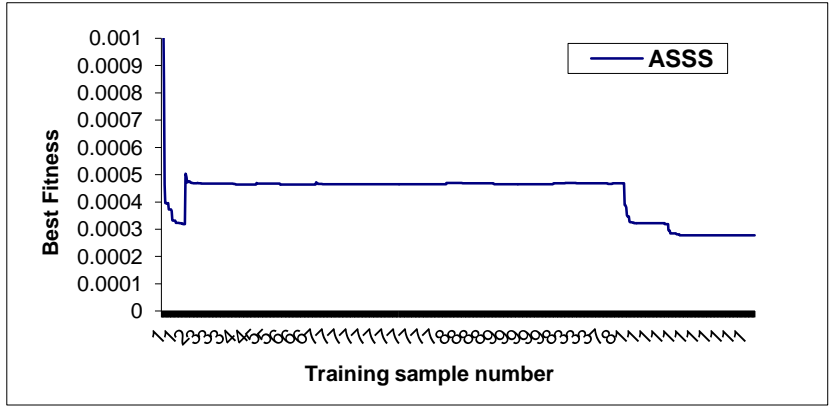

**Figure 4:** Example of curve fitness of the best individuals generated by ASSS method for TS samples

### b. Adaptive-Random training-Subset Selection method (ARSS):

The ARSS method uses the same procedure as the ASSS method, except that the initial weights are generated randomly at the start of running, rather than initialized with a constant: For t=0, $W(S_i) = \tilde{P}_i, \tilde{P}_i \in [0,1], 1 \leq i \leq k.$ Then, for the few first generations, samples are selected using the *RSS* method (subsection 4.2). After, the selection of samples is made using the adaptive approach based on the re-ordering procedure (equation (4)). Figure 5 illustrates an example of the best fitness (MSE) curve along evolution using ARSS method.

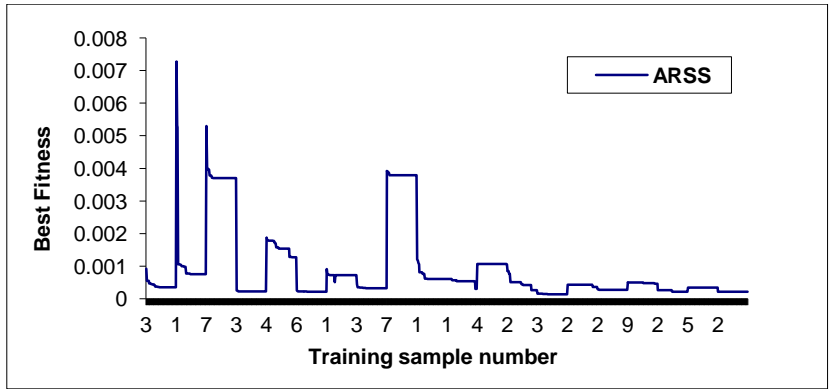

**Figure 5:** Example of curve fitness of the best individuals generated by ARSS method for MTM classes

## V. FINDINGS AND RESULTS ANALYSIS

### 5.1. Experiments

The experiments were performed in two major phases: static subset selection experiments (phase 1) and dynamic subset selection experiments (phase 2).

Phase 1: First, the genetic program is trained separately on each of the first nine TS sub-samples ($S_1,\ldots, S_9$) using ten different seeds and is tested on the subset data from the immediately following date ($S_2,\ldots, S_{10}$). Second, using the same genetic parameters and the random seeds applied on TS data, GP is trained separately on each of the first nine MTM sub-classes ($C_1^L,\ldots, C_9^L$) and is tested on the second nine MTM sub-classes ($C_1^T,\ldots, C_9^T$). Actually, each sub-sample is independently evolvable by GP and the best individual generated from each sub-sample is selected.

Phase 2: First, the genetic program is trained on the first nine TS sub-samples simultaneously ($S_1,\ldots, S_9$) using ten different seeds and it is tested only on the tenth sub-sample data ($S_{10}$). Second, GP is trained on the first nine MTM sub-classes simultaneously ($C_1^L,\ldots, C_9^L$), and it is tested on the second nine MTM sub classes regrouped in one test sample data ($C_1^T + C_2^T \ldots + C_9^T$). Third, GP is trained on both the nine TS sub-samples and the nine MTM sub-classes simultaneously ($S_1, \ldots, S_9$; $C_1^L, \ldots, C_9^L$), and it is tested on one test sample data composed of the TS and MTM test data ($S_{10} + C_1^T + C_2^T \ldots + C_9^T$).

Table 3 summarizes the training and test data samples used for static and dynamic training-subset selection methods respectively.

| Subset Selection | Learning data sample | Test data sample |
| --- | --- | --- |
| Static Subset Selection | 1. $S_i \in$ TS samples ($S_1, \ldots, S_9$) *(1 subset for a run)* | The successive TS sample $S_j$, j=i+1 |
| | 2. $C_i^L \in$ MTM training samples ($C_1^L, \ldots, C_9^L$) *(1 subset for a run)* | The corresponding MTM test sample $C_i^T$ |
| Dynamic Subset Selection (RSS/SSS/ASSS/ARSS) | 1. TS samples $S_1, \ldots, S_9$ *(9 subsets for a run)* | The last subset in TS samples set ($S_{10}$) |
| | 2. MTM training samples $C_1^L, \ldots, C_9^L$ *(9 subsets for a run)* | The nine MTM test samples ($C_1^T + C_2^T \ldots + C_9^T$) |
| | 3. TS samples + MTM samples ($S_1, \ldots, S_9, C_1^L, \ldots, C_9^L$) *(18 subsets for a run)* | The last TS sample with the nine MTM test samples ($S_{10} + C_1^T + C_2^T \ldots + C_9^T$) |

**Table 3**: Definition of training and test data samples for static and dynamic training-subset selection methods.

For each case, the best individual (tree function) is selected according to the MSE's fitness measure. Selected models are then analyzed and compared with each other according to different measures as described in the following.

### 5.2. GP model selection

Results analysis focuses on the comparison between GP solutions given by static and dynamic training-subset selection methods in terms of their ability to forecast implied volatility. GP models subject to the comparative study are selected as follows.

First, selection of the best generated GP volatility model, relative to each training set, for time series (TS), moneyness-time to maturity (MTM), and both TS and MTM classifications, is made according to the training and test MSE. For static training-subset selection method, nine generated GP volatility models ($M_1S_1\ldots M_9S_9$) are selected for TS samples and similarly, nine generated GP volatility models ($M_1C_1\ldots M_9C_9$) are selected for MTM classes.

Second, for dynamic training-subset selection methods (RSS, SSS, ASSS and ARSS), four generated GP volatility models are selected for TS classification (MSR, MSS, MSAS and MSAR). Similarly, four generated GP volatility models are selected for MTM classification (MCR, MCS, MCAS and MCAR) and four generated GP volatility models are selected for global classification using both TS and MTM classes (MGR, MGS, MGAS and MGAR). The following table summarizes the 30 volatility models selected for the comparative study.

| Subset Selection | Learning data | GP volatility models | | | |
|---|---|---|---|---|---|
| Static Subset Selection | TS samples ($S_1, \ldots, S_9$) | $M_1S_1\ldots M_9S_9$ (TS models) | | | |
| | MTM classes ($C_1^L, \ldots, C_9^L$) | $M_1C_1\ldots M_9C_9$ (MTM models) | | | |
| Dynamic Subset Selection | | **RSS** | **SSS** | **ASSS** | **ARSS** |
| | TS series ($S_1, \ldots, S_9$) | MSR | MSS | MSAS | MSAR |
| | MTM classes ($C_1^L, \ldots, C_9^L$) | MCR | MCS | MCAS | MCAR |
| | TS series + MTM classes ($S_1, \ldots, S_9, C_1^L, \ldots, C_9^L$) | MGR | MGS | MGAS | MGAR |

**Table 4**: Definition of GP generated models given by static and dynamic training-subset selection methods.

### 5.3. Results analysis

To assess the accuracy of the training-subset selection methods applied as well as the generated GP volatility models selected, two measures are used. First, the **MSE Total** is computed using the same formula as the basic MSE (equation 3) but according to the enlarged data sample. The best model is that provides the smallest forecasting error. MSE Total aims to measure the generalization ability of the GP generated models. Second, **the number of non-fitted observations (NFO)** for a given data sample is used to describe how well a model fits the sample observations; otherwise it measures the model's ability to solve all fitness cases in the corresponding subset. An observation is supposed to be well fitted if the corresponding error (The absolute difference between target and forecasted output volatility) is less than 0.1 (value determined according to the experimental results). NFO aims to compare the final solutions in terms of adaptation level to all input fitness cases.

Figure 6 describes the performance of the best generated GP volatility models, using static and dynamic training-subset selection methods, according to the MSE total for all data samples.

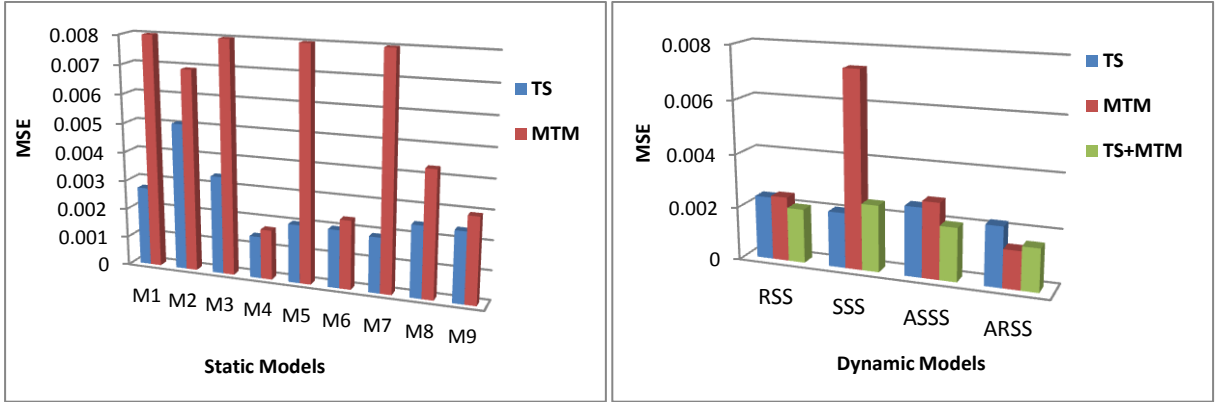

Figure 6 (a)          Figure 6 (b)

**Figure 6**: Performance of the generated GP volatility models, using static and dynamic training-subset selection methods, according to the MSE total for TS samples ($S_1\ldots S_9$), MTM classes ($C_1^L\ldots C_9^L$) and both TS and MTM samples ($S_1\ldots S_9, C_1^L\ldots C_9^L$).

Figure 6 (a) presents the MSE total relative to the 18 generated GP volatility models, using static training-subset selection method, selected for TS samples and for MTM classes. Some extreme MSE values for MTM data are not shown in this figure. As shown, the performance of the static models is not uniform. Total errors are higher for the MTM classes than for the *TS* samples. Indeed, the MSE exceed 1 with some fitness cases of MTM classes, and it doesn't reach 0.006 for all *TS* sample cases. Thus, it seems that TS models are more general than MTM models. The difference in accuracy between GP applied on TS samples and GP applied on MTM classes is very striking. With MTM classes, GP was unable to find satisfactory models with high forecasting ability, which might be caused by insufficient search intensity. Furthermore, Figure 6 (a) shows that the generated GP models

M4S4 and M4C4 have the smallest MSE in enlarged sample, for the TS and for the MTM classes respectively. They seem to be more accurate in forecasting implied volatility than the other models.

Figure 6 (b) illustrates the MSE relative to the 12 generated GP volatility models, using dynamic training-subset selection methods (RSS, SSS, ASSS and ARSS), selected for TS samples, for MTM classes and for global classification using both TS and MTM classes. It appears throughout Figure 6 (b) that the generated GP volatility models, relative to each dynamic subset selection method, are performing on the enlarged sample and present forecasting errors which are small and much close. The MSE relative to these models don't reach 0.003 for all sampling scheme data, except the MCS model generated using the SSS method for MTM classes. Figure 6 (b) shows that, for the TS samples, the MSS model generated using the SSS method has the smallest MSE on the enlarged sample. For the MTM classes, the MCAR model generated using ARSS method outperforms the other models generated using the other methods. For both TS and MTM data, the MGAR model generated using ARSS method presents the highest accuracy in enlarge sample. Overall, the best forecasting's performance is achieved by the ARSS method. This can be explained by the fact that this method permits to generate more general models adaptive to all sample data.

Comparison between static (Figure 6 (a)) and dynamic training- subset selection methods (Figure 6 (b)) in terms of MSE reveals that the quality of the generated GP models has been improved with the dynamic training, particularly for MTM classes. The amplitude of forecasting errors relative to MTM classes is lower for the models generated using dynamic training- subset selection methods than for the models generated using static training- subset selection method.

In order to demonstrate more explicitly the improvements accomplished by the dynamic subset selection, the percentage of non fitted observations (NFO) in the whole learning sample is computed for the best generated GP models using static and dynamic training-subset selection methods. Results are illustrated in Figure 7 for TS samples, MTM classes and both TS and MTM samples.

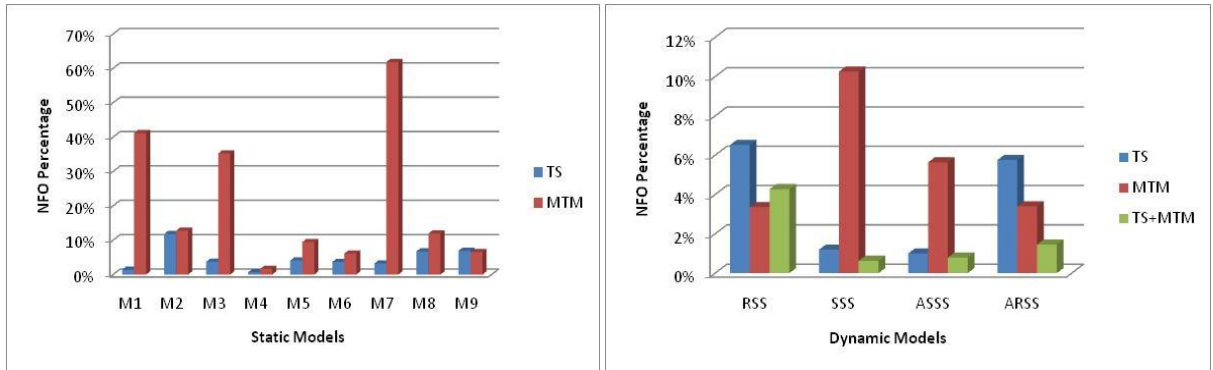

Figure 7 (a)  Figure 7 (b)

**Figure 7:** Performance of the generated GP volatility models, using static and dynamic training- subset selection methods, according to the percentage of non fitted observations (NFO) for TS samples ($S_1…S_9$), MTM classes ($C_1^L…C_9^L$) and both TS and MTM data ($S_1…S_9, C_1^L…C_9^L$).

Figure 7 (a) describes the performance of the 18 generated GP volatility models relative to TS samples and MTM classes, using static training- subset selection method, according to the percentage of NFO.

It appears throughout Figure 7 (a) that the TS models are best-fit patterns compared to MTM models. In fact, the percentages of NFO given by TS models are markedly lower than those achieved by MTM models for the enlarged sample. The NFO percentages given by TS models do not exceed 12%, but they reach 62% with MTM models. Indeed, training is more homogeneous with TS samples than with MTM classes. This leads us to affirm that the static subset selection depends on the sampling scheme. Otherwise, it's important to note that each model (from TS models or MTM models) fits well to the data on which is trained. Actually, the generated GP volatility models $M_4S_4$ and $M_4C_4$ present the lower percentages of NFO for TS and MTM samples respectively.

Results show that the gap between the models' performances is more remarkable for MTM classes. This can be explained by the fact that the majority of models fit well their training classes but not other out-of samples. GP is most generally not efficient when the training data pattern is different from the one relative to out-of-sample data (Chen (2007)).

Figure 7 (b) describes the performance of the 12 generated GP volatility models relative to TS samples, MTM classes and both TS and MTM classes, using dynamic training-subset selection methods, according to the percentage of NFO in the enlarged sample. This percentage varies from 0.63% (MGS) to 4.27% (MGR) for global data training, and from 1 % (MSAS) to 6.52 % (MSR) for TS training. Corresponding solutions could be considered as robust models with high forecasting accuracy. For MTM training, the NFO percentage is relatively high and varies from 3.36% (MCR) to 10.25% (MCS).

It could be observed that the dynamic methods (SSS, ASSS, ARSS) have the highest performance with global classification data (both TS and MTM) than with TS samples or MTM classes. In fact, the GP volatility models generated using these methods have NFO percentages lower for global classification (MGS, MGAS and MGAR) than for TS samples (MSS, MSAS and MSAR) and MTM classes (MCS, MCAS and MCAR) respectively. The RSS method gives high training quality for the MTM classes. This can be attributed to the randomness of the learning sample order which allows a better adaptation of the population to the environmental changes. Although the SSS method provides the best accuracy with global classification (NFO =0.63%) and less to TS samples (NFO=1.2%), it was unable to provide high performance for MTM classes (NFO=10.25%). This can be explained by the fact that the sample selection scheme with the SSS method is steady and unchanged along evolution.

Comparison between static (Figure 7 (a)) and dynamic training- subset selection methods (Figure 7 (b)) reveals that the percentages of NFO are in most cases lower for the models generated using dynamic training- subset selection methods than for the models generated using static training-subset selection method. The NFO percentage is reduced for most samples, in particular the MTM classes when using dynamic training-subset selection methods.

Overall, according to the MSE Total and the NFO percentage, the generated GP models using dynamic training- subset selection methods exhibit a very high accuracy relative to that using static training-subset selection method. This observation is confirmed by the measures in Table 5, which illustrates the average of MSE total and NFO percentage for all the models obtained with static and dynamic subset selection for each set of learning data samples (TS series, MTM classes and both TS series and MTM classes). The dynamic subset selection was able to achieve the desired goal and improve the GP research process in order to fit better the whole learning data. It presents the smallest averages of MSE Total and NFO percentage. Otherwise, we can note that the diversity of the input samples in the case of dynamic selection applied on both TS and MTM samples makes GP more robust in supervised learning and so as the generated forecasting models outperform all other models.

| Measures \ Methods | Static Subset Selection | | Dynamic Subset Selection | | |
|---|---|---|---|---|---|
| | *TS* series | *MTM* classes | *TS* series | *TM* classes | *TS* series + *MTM* classes |
| Average of MSE Total | 0.002599 (0.064383) | 0.416320 (56.067) | 0.002372 (0.003894) | 0.003600 (0.126530) | **0.002033 (0.003506)** |
| Average of NFO percentage | 4.27% | 20.29% | 3.41% | 5.67% | **1.77%** |

**Table 5:** Average of MSE Total and NFO percentage for static and dynamic subset selection applied to the different learning data sets. The numbers in parentheses are the standard deviation corresponding to MSE values of all observations in each sample set.

The last step of the present work is the selection of the most accurate models from the 30 GP generated models as listed in table 4.

Based on the MSE total and the percentage of NFO as performance criteria, the generated GP volatility models $M_4S_4$ and $M_4C_4$ are selected for static training-subset selection method. Similarly, the generated GP volatility models MSS, MCAR and MGAR are selected for dynamic training-subset selection method. Table 6 reports the performance of these selected models.

| Models | MSE total | Percentage of NFO |
|--------|-----------|-------------------|
| **M4S4** | **0.001444 (0.002727)** | **0.67%** |
| M4C4 | 0.001710 (0.004624) | 1.66% |
| MSS | 0.002076 (0.004044) | 1.20% |
| **MCAR** | **0.001424 (0.003527)** | **3.40%** |
| **MGAR** | **0.001599 (0.003590)** | **1.45%** |

**Table 6:** Selection of the best generated GP volatility models, using static and dynamic training-subset selection methods, in terms of MSE total and percentage of NFO.

Results show that according to MSE total and NFO percentage, the TS model $M_4S_4$ seems to be more performing than the MTM model $M_4C_4$ for static training-subset selection method. It presents the lowest MSE and percentage of NFO.

According to MSE total, the MCAR and MGAR models generated using the ARSS method seem to outperform the MSS model generated using the SSS method. Although the latter has the lowest percentage of NFO, it seems to be less performing than the MCAR and MGAR models. This can be explained by two points. First, the time series model MSS presents the highest MSE relative to the other models. Second, the total of error values computed for the non fitted observations are higher for MSS than the other models even it presents the lowest percentage of NFO.

Comparison between models reveals that the best models generated respectively by static ($M_4S_4$) and dynamic selection methods (MCAR and MGAR) present small and very close total MSE values. While the generated GP volatility models $M_4S_4$ and MCAR have total MSE smaller than the MGAR model, the latest seems to be more accurate in forecasting implied volatility than the other models. This can be explained by the fact that, on one hand, the difference between forecasting errors is small, and on the other hand, the MGAR model is more general than MCAR and $M_4S_4$ models because it is adaptive to all time series and moneyness-time to maturity classes simultaneously. According to the percentage of NFO, the MGAR model presents a percentage of NFO relatively higher than the $M_4S_4$ model, trained only on time series data, and relatively smaller than the MCAR model, trained only on MTM classes.

The decoding of these models yields the following GP volatility forecasting formulas:

M4S4: $$\sigma_{GP} = \exp\left[\left(\ln\left(\Phi\left(\frac{C}{K}\right)\right)\right) * \sqrt{\tau - 2*\frac{C}{K} + \frac{S}{K}} - \cos\left(\frac{C}{K}\right)\right]$$

MCAR: $$\sigma_{GP} = \frac{\dfrac{\tau * \sqrt{\dfrac{C}{K}}}{\ln\left(\dfrac{S}{K} + \tau\right)}}{\left(\dfrac{S}{K}\right)^2 * \left(\dfrac{S}{K} + \tau\right)}$$

MGAR: $$\sigma_{GP} = \sqrt{\frac{\dfrac{C}{K}}{\left(\dfrac{S}{K}\right)^6 + \left(\dfrac{S}{K}\right)^5 * \tau}}$$

## VI. CONCLUSION

This article presents a Genetic Programming based technique to generate implied volatility forecasting models from S&P500 index options. We have demonstrated that the accuracy of the generated models depends on the training sample, especially when learning with moneyness- time to maturity classes. To reduce the gap between in-sample fit and out-of-sample performance, we introduced the dynamic training which aims to enlarge the training set to the whole input data. Four techniques of dynamic subset selection are proposed: Random training-Subset Selection (RSS) where samples are selected in random way, Sequential training-Subset Selection (SSS) where samples are selected in a regular way, Adaptive-Sequential training-Subset Selection (ASSS) and Adaptive-Random training-Subset Selection (ARSS) which use a training instance weight to enhance learning on the "difficult" fitness cases. These techniques are applied on the time series samples and on moneyness- time to maturity classes, and compared to the static training subset selection method using a single sample for the learning process.

Experiments indicate that using the dynamic training with GP yields better results than applying the static training, especially when learning on time series and moneyness- time to maturity samples simultaneously. Otherwise, based on the MSE total and the percentage of NFO as performance criteria, three generated GP volatility models are selected: M4S4 generated using the static training-subset selection method**,** MCAR generated using the ARSS method applied on moneyness- time to maturity classes and MGAR generated using the ARSS method applied on times series and moneyness- time to maturity classes regrouped. However, the MGAR seems to be more accurate in forecasting implied volatility than MCAR and M4S4 models. This means that the MGAR model is more general than MCAR and M4S4 models because it is adaptable to all time series and moneyness-time to maturity classes simultaneously.

Our results revealed some interesting issues for further investigation. First, the dynamic training GP can be used to forecast implied volatility of other models than BS model, notably stochastic volatility models and models with jump. Second, this work can be reexamined using data from individual stock options, American style index options, options on futures, currency and commodity options. Third, the performance of the generated GP volatility models can be measured in terms of trading and hedging. Finally, the GP approach can be applied to extract risk-neutral densities, which provide valuable information about market expectations. We believe that these extensions are of interest for application and will be the object of our future works.

## Appendix: Symbols and abbreviations

| Genetic Programming Symbols | Financial symbols |
|---|---|
| GP : Genetic Programming<br>GA: Genetic algorithms<br>NFO: non fitted observations<br>MSE: mean squared error<br><br>Dynamic subset selection Symbols<br><br>SSS: Sequential Subset Selection method<br>RSS: Random Subset Selection method<br>ASS: Adaptive Subset Selection method<br>ASSS: Adaptive-Sequential Subset Selection method<br>ARSS: Adaptive-Random Subset Selection method | C:  call option price<br>K: strike price<br>S: the index price<br>$\tau$ : time to maturity<br>ITM: in-the-money<br>OTM : out-of-the-money<br>ATM : at-the-money<br>ST: short-term<br>MT: medium-term<br>LT: long-term<br>TS: times series<br>MTM: moneyness-time to maturity |

## References

◊ Abdelmalek, W., Ben Hamida, S. and Abid, F. (2009), "Selecting the Best Forecasting- implied Volatility Model using Genetic Programming", *Journal of Applied Mathematics and Decision Sciences*, Hindawi Publishing Corporation. www.hindawi.com/journals/jamds/2009/179230.html

◊ Baillie, R. T., Bollerslev, T., and Mikkelsen, H. O. (1996), "Fractionally integrated generalized autoregressive conditional heteroskedasticity", *Journal of Econometrics*, vol.74, n°1, pp. 3-30.

- Banzhaf, W., Nordin, P., Keller, R. E., and Francone, F. D. (1998), Genetic Programming. An Introduction. Morgan Kaufmann, San Francisco, CA.

- Black, F. and Scholes, M. (1973), "The Pricing of Options and Corporate Liabilities", *Journal of Political Economy*, vol. 81, n°3, May-June, pp. 637-654.

- Blair, B.J., Poon, S. and Taylor, S.J. (2001), "Forecasting S&P100 Volatility: The Incremental Information Content of Implied Volatilities and High Frequency Index Returns", *Journal of Econometrics*, vol. 105, pp. 5 -26.

- Bollerslev, T. (1986), "Generalized autoregressive conditional heteroscedasticity", *Journal of Econometrics*, vol. 31, pp. 307-328.

- Cai, W., Pacheco-Vega, A., Sen, M. and Yang, K.T (2006), "Heat Transfer Correlations by Symbolic Regression", *International Journal of Heat and Mass Transfer*, vol. 49, pp. 4352-4359.

- Chen, S.H. and Yeh, C.H. (1997), "Using Genetic Programming to Model Volatility in Financial Time Series", *Genetic Programming 1997: Proceedings of the Second Annual Conference*, Morgan Kaufmann Publishers, pp. 288-306.

- Chen, S.H., Kuo, T.Z. and Koi, K.M. (2007), Genetic Programming and financial trading: How much about "what we know". Handbook of financial Engineering.

- Chiras, D. and Manaster, S. (1978),"The Information Content of Option Prices and a Test of Market Efficiency", *Journal of Financial Economics*, vol. 6, pp. 213-234.

- Corrado, C.J. and Miller, T.W. (2005), "The Forecast Quality of CBOE Implied Volatility Indexes", *The Journal of Futures Markets*, vol. 25, Issue 4, pp. 339–373.

- EO library :  Evolving Object library, an Evolutionary Computation Framework coded in C++ language (http://eodev.sourceforge.net/)

- Fleming, J. (1993), "The Rationality of Market Volatility Forecasts Implied by S&P 100 Index Option Prices", *Working Paper*, Fuqua School of Business, Duke University.

- Gathercole, C. and Ross, P. (1994), "Dynamic Training Subset Selection for Supervised Learning In Genetic Programming", *Parallel Problem Solving from Nature III*, vol. 866 of LNCS, pp. 312-321.

- Gathercole, C. and Ross, P. (1997), "Small populations over many generations can beat large populations over few generations in genetic programming", Genetic *Programming 1997: Proceedings of the Second Annual Conference*, pages 111-118, Stanford University, CA, USA. Morgan Kaufmann.

- Gathercole, C. (1998), "An Investigation of Supervised Learning in Genetic Programming", PhD thesis, University of Edinburgh.

- Gilli, M. and Schumann, E. (2010), "Heuristic Optimisation in Financial Modelling", *Annals of Operations Research*, Available at SSRN: http://ssrn.com/abstract=1277114.

- Gustafson, M., Burke, E.K. and Krasnogor (2005), "On Improving Genetic Programming for Symbolic Regression", *in Proceedings of the IEEE Congress on Evolutionary Computation*, vol. 1, pp. 912-919.

- Harvey, C.R. and Whaley, R.E. (1991), "S&P 100 Index Option Volatility", *Journal of Finance*, vol. 46, pp. 1551-1561.

- Harvey, C.R. and Whaley, R.E. (1992), "Market Volatility Prediction and the Efficiency of the S&P100 Index Option Market", *Journal of Financial Economics*, vol. 31, pp. 43-73.

- Ito, T., Iba, H. and Kimura, M. (1996), "Robustness of Robot Programs Generated by Genetic Programming", *in Genetic Programming 96*, Mit Press.

◊ Keijzer, M. (2004), "Scaled Symbolic Regression", *Genetic Programming and Evolvable Machines*, vol. 5, n°3, pp. 259-269.

◊ Koza, J.R. (1992), Genetic Programming: on the Programming of Computers by means of Natural Selection. Cambridge, Massachusetts: the MIT Press.

◊ Latané, H.A. and Rendleman, R.J. (1976), "Standard Deviations of Stock Price Ratios Implied in Option Prices", *Journal of Finance*, vol. 31, pp. 369-381.

◊ Lasarczyk C. W. G., Dittrich P. and Banzhaf W. (2004), "Dynamic subset selection based on a fitness case topology", *Evolutionary Computation* 12(2):223-242.

◊ Lew, T.L., Spencer, A.B., Scarpa, F. and Worden, K. (2006), "Identification of Response Surface Models Using Genetic Programming", *Mechanical Systems and Signal Processing*, vol. 20, n°8, pp. 1819-1831.

◊ Ma, I., Wong, T. and Sanker, T. (2006), "An Engineering Approach to Forecast Volatility of Financial Indices", *International Journal of Computational Intelligence*, vol. 3, n°1, pp. 23-35.

◊ Ma, I., Wong, T. and Sanker, T. (2007), "Volatility Forecasting using Time Series Data Mining and Evolutionary Computation Techniques", *Proceedings of the 9$^{th}$ Annual Conference on Genetic and Evolutionary Computation (GECCO 07)*, England, ACM Press.

◊ McKay, B., Willis, M.J. and Barton, G.W. (1995), "Using a Tree Structural Genetic Algorithm to Perform Symbolic Regression", *First International Conference on Genetic Algorithms in Engineering Systems: Innovations and Applications (GALESIA)*, vol. 414, pp. 487-492.

◊ Merton, R.C. (1973), "Theory of Rational Option Pricing", *Bell Journal of Economics and Management Science*, vol. 4, pp. 141-183.

◊ Müller, U.A., Dacorogna, M.M., Davé, R.D., Olsen, R.B., Pictet O. V., and von Weizsäcker, J.E. (1997), "Volatilities of different time resolutions-analyzing the dynamics of market components", *Journal of Empirical Finance*, vol.4, n° 2-3, pp. 213-239.

◊ Neely, C.J. and Weller, P.A. (2002), "Using a Genetic Program to Predict Exchange Rate Volatility", *in Genetic Algorithms and Genetic Programming in Computational Finance*, Chapter 13, Edited by Chen S.H, Kluwer Academic Publishers, pp. 263-279.

◊ Nordin, P. and Banzhaf, W. (1997), "An on-line method to evolve behavior and to control a miniature robot in real time with genetic programming", *Adaptive Behaviour*, vol. 5, n°2, pp.107-140.

◊ Povinelli, R.J. (1999), "Times Series Data Mining: Identifying Temporal Patterns for Characterization and Prediction of Time Series Events", Marquette University, Ph. D.

◊ Schwefel H.P. (1995), Numerical Optimization of Computer Models, John Wiley & Sons, New York, 1981; 2nd edition.

◊ Yin, Z., Brabazon, A., O'Sullivan, C. and O'Neil, M. (2007), "Genetic Programming for Dynamic Environments", *Proceedings of the International Multiconference on Computer Science and Information Technology*, pp. 437-446.

◊ Zumbach, G., Pictet, O.V. and Masutti, O. (2001). "Genetic Programming with Syntactic Restrictions Applied to Financial Volatility Forecasting", *Technical Report GOZ.2000-07-28*, Olsen & Associates Research Institute, pp. 1-22.